\documentclass[9pt,twocolumn,twoside]{article}
\usepackage{fancyhdr}
\usepackage{authblk}
\usepackage{abstract}
\renewcommand{\abstractname}{} 

\usepackage{graphicx}
\usepackage{textcomp}
\usepackage[margin=0.5in]{geometry}
\usepackage{enumitem}
\usepackage{epstopdf}
\title{\textbf{Speckle-based hyperspectral imaging combining multiple scattering and compressive sensing in nanowire mats}}

\author[1]{Rebecca French}
\author[2]{Sylvain Gigan}
\author[1,*]{Otto L. Muskens}

\affil[1]{\small{\textit{Faculty of Physical and Applied Sciences, University of Southampton, Highfield, Southampton SO17 1BJ, UK}}}
\affil[2]{\small{\textit{Laboratoire Kastler Brossel, ENS-PSL Research University,CNRS, UPMC-Sorbonne Universit\'{e}s, Coll\`{e}ge de France, 24 rue Lhomond, 75005 Paris, France}}}

\affil[*]{Corresponding author: O.Muskens@soton.ac.uk}

\begin{document}
	\twocolumn[
	\begin{@twocolumnfalse}
		\date{}
		
	\maketitle

\begin{abstract}
	\normalsize Encoding of spectral information onto monochrome imaging cameras is of interest for wavelength multiplexing and hyperspectral imaging applications. Here, the complex spatio-spectral response of a disordered material is used to demonstrate retrieval of a number of discrete wavelengths over a wide spectral range. Strong, diffuse light scattering in a semiconductor nanowire mat is used to achieve a highly compact spectrometer of micrometer thickness, transforming different wavelengths into distinct speckle patterns with nanometer sensitivity. Spatial multiplexing is achieved through the use of a microlens array, allowing simultaneous imaging of many speckles, ultimately limited by the size of the diffuse spot area. The performance of different information retrieval algorithms is compared. A compressive sensing algorithm exhibits efficient reconstruction capability in noisy environments and with only a few measurements.\\
	
		\footnotesize{\textcopyright}  \small   2017 Optical Society of America. One print or electronic copy may be made for personal use only. Systematic reproduction and distribution, duplication of any material in this paper for a fee or for commercial purposes, or modifications of the content of this paper are prohibited.
	\vspace{1cm}
	\renewcommand{\abstractname}{\vspace{-\baselineskip}}

\end{abstract}

\end{@twocolumnfalse}
]

	\begin{figure*}[ht]
		\centering
		\includegraphics[width=0.85\linewidth]{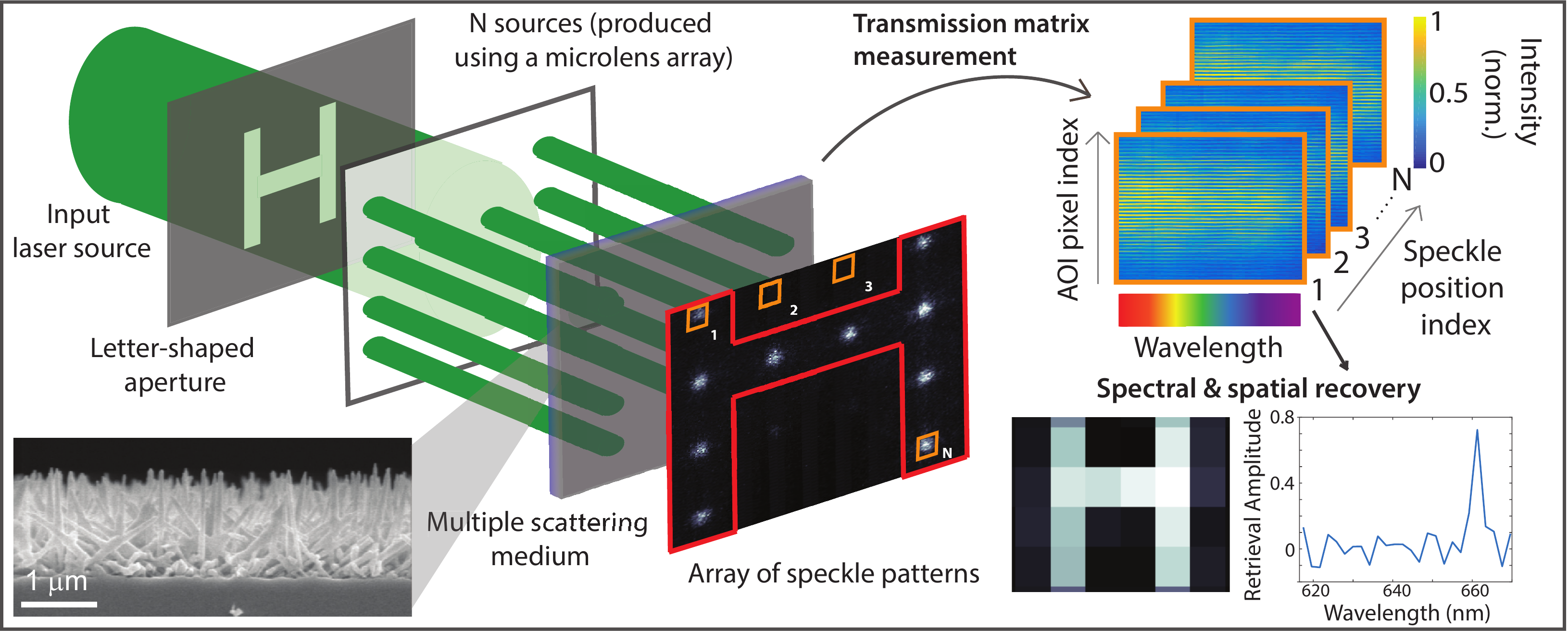}
		\caption{Scheme of speckle-based hyperspectral camera. A microlens array separates the input light into a grid of N sources. The light travels through the multiple scattering medium (SEM in inset) producing an array of speckle patterns which are imaged onto a camera. Mapping the spectral patterns at each spatial position using a tunable source results in the STM. Reconstruction of a letter: An input beam in the shape of an `H' is sent through the imaging system and forms an intensity pattern. The pre-calibrated STM is used to reconstruct an image of the original incoming beam and extract its corresponding wavelength. }
		
		\label{fig:Experiment}
	\end{figure*}
	With the advancement of modern camera technology, sacrificing spatial resolution to obtain additional functionality is an attractive way to extract more information from a single exposure. In particular, the incorporation of wavelength information into a spatial image is of interest as it enables the measurement of multispectral datasets. Several solutions have been demonstrated based on spectral encoding using diffractive or refractive elements \cite{Orth:15,Gehm:07,Gao:10}.  While impressive results have been achieved, the trade-off between the accessible spectral range and the resolution in such linearly dispersive systems poses limitations on their performance in specific applications. Especially in cases where intensity is concentrated in a few narrow spectral features over a wide spectral range, conventional dispersive techniques will not make optimal use of the available pixel space.  Alternative methods of mapping spectral content in a way that distributes energy more equally over all available pixels are of interest, as these can exploit the full dynamic range of the imaging system. A recent example showed that diffraction from deterministically designed structures can be used to obtain multispectral images \cite{Wang2015}.
	
	In recent years, the utilization of multiple scattering in imaging and sensing has seen an increase in interest. The emergence of wavefront shaping and transmission matrix techniques has raised the awareness that random multiple scattering can provide powerful tools for manipulating information in the spatial domain. These techniques have enabled the focusing of light and transmission of images through multiple scattering media \cite{Popoff2010,Popoff20102, Vellekoop2007}. It has also been suggested that multiple scattering could be combined with compressive sensing (CS) to enable more efficient imaging and spectroscopy \cite{Liutkus2014,Shin:16}. In the spectral domain, it is well known that the output intensity pattern, or speckle of light, after traveling through a multiple scattering medium is frequency-dependent \cite{deBoer1992, Feng1988, Genack1987}. A transmission matrix approach can be used to store the different spectral fingerprints for a desired frequency range in order to characterize arbitrary wavelengths \cite{Kohlgraf2010, Xu2010, Hang2010}. A compact spectrometer has been demonstrated by characterizing the frequency channels of a multimode fiber, where the spectrometer resolution is dependent on the length of the fiber \cite{Redding2012, Redding2014}. Other recent examples of speckle-based spectrometers have utilized the memory-effect, principal-component analysis of spectral intensity patterns, or a disordered photonic crystal to achieve a compact spectrometer \cite{Chakrabarti2015, Mazilu2014,Redding20132}.
	
	Here, we demonstrate a highly multiplexed transmission matrix-based spectrometer to achieve a hyperspectral imaging system. A 1.7 \textmu m thin layer of strongly scattering gallium phosphide nanowires is used to provide a uniform, highly-dispersive scattering medium of only a few micrometers in thickness. A scanning electron microscopy (SEM) image of the cross-section through the nanowire mat is shown in Figure~\ref{fig:Experiment}. The nanowires were grown using metal-organic vapor phase epitaxy described in detail elsewhere \cite{Muskens2009, Muskens2008}. This method allows fabrication of highly uniform layers over wafer-scale size. Multiple scattering in the nanowire mat results in a transport mean free path of $\ell=300$~nm, yielding a transmission $T \simeq \ell/L$ of around 17\%. The spectral correlation width is given by the inverse Thouless time $\tau_D^{-1}=D/L^2=4.8$~THz (corresponding to 7~nm at 650~nm wavelength), where $D = 14 \pm 1$~m$^2$/s is the measured diffusion constant of light inside the nanowire mat \cite{Strudley2013}. The nanowire mat contains around $100$ independent transmission channels within the illumination area \cite{Strudley2013}. This number is sufficient to encode spectral information but is also small enough to yield good stability. Furthermore, the thin medium allows a potentially high spatial resolution as the diffuse spreading of the spots is small.
	
	Figure~\ref{fig:Experiment} shows the experimental design for the characterization of spectrally- and spatially-dependent speckle patterns. Light from a white light supercontinuum laser (Fianium, SC-400-2) is spectrally filtered using a monochromator consisting of a 600~lines/mm grating and a 10~\textmu m single-mode optical fiber, resulting in a tunable source with spectral resolution of 0.5~nm and several mW average power. The collimated light from the single-mode fiber is used as an illumination source for imaging. The objects in this study are aperture masks of around 1.5~mm in size. After transmission through the object, the light is focused onto the scattering medium by using a microlens array with a pitch of 300~\textmu m, producing a grid of around $N=100$ input spatial positions. Each microlens produces a diffraction-limited spot of 10~\textmu m.
	
	Intensity patterns produced by multiple scattering at the exit plane of the scattering medium are imaged onto the camera using a 1:1 imaging system consisting of a pair of 3~cm focal length lenses. The detection camera (AVT Guppy) is a 12-bit, 5 MPixel monochrome CMOS array of 2.4~\textmu m pixel size. The imaging magnification was chosen to yield a speckle size close to the size of the camera pixels, thus optimizing the information density. While in future applications the nanowires could, with appropriate spacers, be positioned directly onto the sensor surface itself, the current configuration offers some flexibility for placing additional components. A linear polarizer was used to increase the speckle contrast.
	
	\begin{figure*}[ht]
		\centering
		\includegraphics[width=0.9\linewidth]{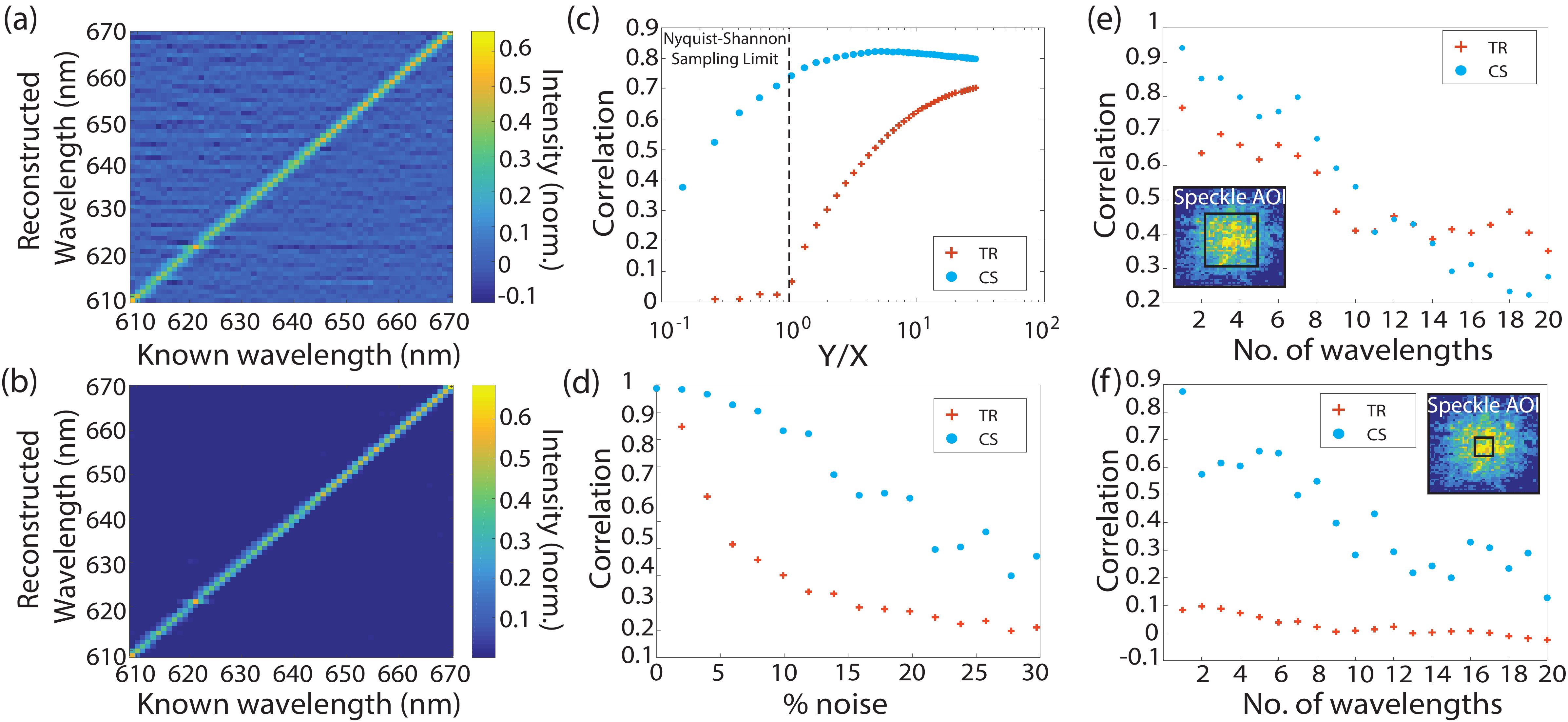}
		\caption{(a,b) A comparison between the known input wavelengths and reconstructed wavelengths using TR and CS, respectively. (c-f) Cross-correlations between known wavelengths and an experimental reconstruction: (c) as the ratio between the number of measurements (or AOI size) Y and number of wavelengths in data set X is varied, averaged over many spatial positions; (d) with increasing noise using different computational methods ($Y/X = 25$), averaged over many spatial positions; (e,f) with an increasing number of discrete wavelengths recovered from one spatial position, using 2 different measurement sizes: above ($Y/X = 7$) and below ($Y/X = 0.8$) the Nyquist-Shannon limit, respectively. Data from (c,e,f) were correlated between two subsequent datasets.}
		\label{fig:CompReconstruct}
	\end{figure*}
	
	The wavelength-dependent spectral intensity transmission matrix (STM) for each position is obtained by selecting the same square area of interest (AOI) from each speckle pattern for wavelengths in the range of 610~nm to 670~nm, with separation of 1~nm. For each spatial coordinate, the 2D speckle images are reshaped into column vectors to create a STM. As illustrated in Figure~\ref{fig:Experiment}, the stack of STMs can be used to reconstruct full spatial and spectral information for a given image, where each spatial position becomes one ``pixel" in our reconstruction. The measured STM remains stable for periods of up to an hour, mainly limited by the optical setup.
	
	The retrieval of spectral information using the STM can be treated as a linear problem $ y = T x $, where $x$ is an input in the system, $y$ is the resulting output signal, and $T$ is the STM. Knowledge of the STM allows us to rearrange this equation to find the original input, and hence to determine the spatial and spectral information of the original input signal. Several mathematical inversion techniques can be used. The method of Tikhonov regularization (TR) is able to account for the experimental noise by suppressing divergences in the singular values below a critical noise parameter \cite{Popoff20102, Tikhonov1963}. TR requires careful adaptation of the noise level and in practice works best for a large AOI, as will be shown below. Figure~\ref{fig:Experiment} shows a reconstruction of the letter 'H' for a single-wavelength input. Here, the individual coordinates of the microlens array were converted to a bitmap image with grayscale representing the retrieval amplitude at the chosen wavelength of 661~nm.
	
	An alternative method of determining the original input in a system is to employ CS in our computational reconstruction; more specifically, using l$_{1}$-minimization. Candes, Romberg, Tao, and Donoho established that a sparse signal can be completely reconstructed in a number of measurements less than the Nyquist-Shannon limit \cite{Candes2006,Donoho2006}. It was suggested that a multiple scattering medium with natural randomness could be employed in a CS device \cite{Liutkus2014}. In our CS reconstruction we used CVX, a package for specifying and solving convex programs \cite{CVX}. Figures~\ref{fig:CompReconstruct} (a) and (b) show the reconstruction of a broad range of wavelengths to compare TR and CS. While both methods are able to accurately reconstruct wavelengths, CS is able to minimize reconstruction noise while TR produces many small values fluctuating around zero. The width of the diagonal on both graphs defines the correlation between  speckles of neighboring wavelengths. The reconstruction amplitudes are slightly higher toward longer wavelengths due to an increase in available laser intensity and correspondingly better signal-to-noise.
	
	Figures~\ref{fig:CompReconstruct} (c-f) show computational experiments in which all the resulting reconstructions were cross-correlated with the original input. Figure~\ref{fig:CompReconstruct} (c) shows the reconstruction quality between two different experimental datasets whilst varying the ratio of the number of available speckles, $Y$, to the number of independent wavelengths, $X$. As the AOI size is reduced from overdetermined ($Y/X>1$) to underdetermined ($Y/X<1$), the fidelity of the results produced by the TR method is seen to decrease. In comparison, CS performs consistently well even for underdetermined systems.
	
	In Figure~\ref{fig:CompReconstruct} (d) the same dataset was used both for the reconstruction and the correlation, however a percentage of artificial randomly-generated exponential noise was added to the measured speckle pattern to investigate the effect of noise on the reconstruction. CS shows improved performance, even under noisy experimental conditions, when compared with TR despite being optimized to account for the increase in noise.
	
	To investigate the ability of our system to reconstruct a full spectrum, CS and TR were both used to recover an increasing number of wavelengths, with uniform separation of the order of the spectral correlation bandwidth of the system, in one acquisition. Figures~\ref{fig:CompReconstruct} (e) and (f) show the correlation between the two experimental datasets for a STM measured above and below the Nyquist-Shannon limit, respectively. Using a large number of measurements (large AOI) allows a reconstruction of up to 10 wavelengths using CS, assuming a correlation of 0.5 or more. In both cases, CS outperforms the TR method. The CS sparsity condition dictates that the method is less advantageous for a large number of wavelengths, and so it appears that TR surpasses CS above 14 wavelengths for a large AOI. All of these measurements provide evidence that a CS technique can dramatically reduce the amount of data required to fully reconstruct a signal.
	
\begin{figure}
		\centering
		\includegraphics[width=\linewidth]{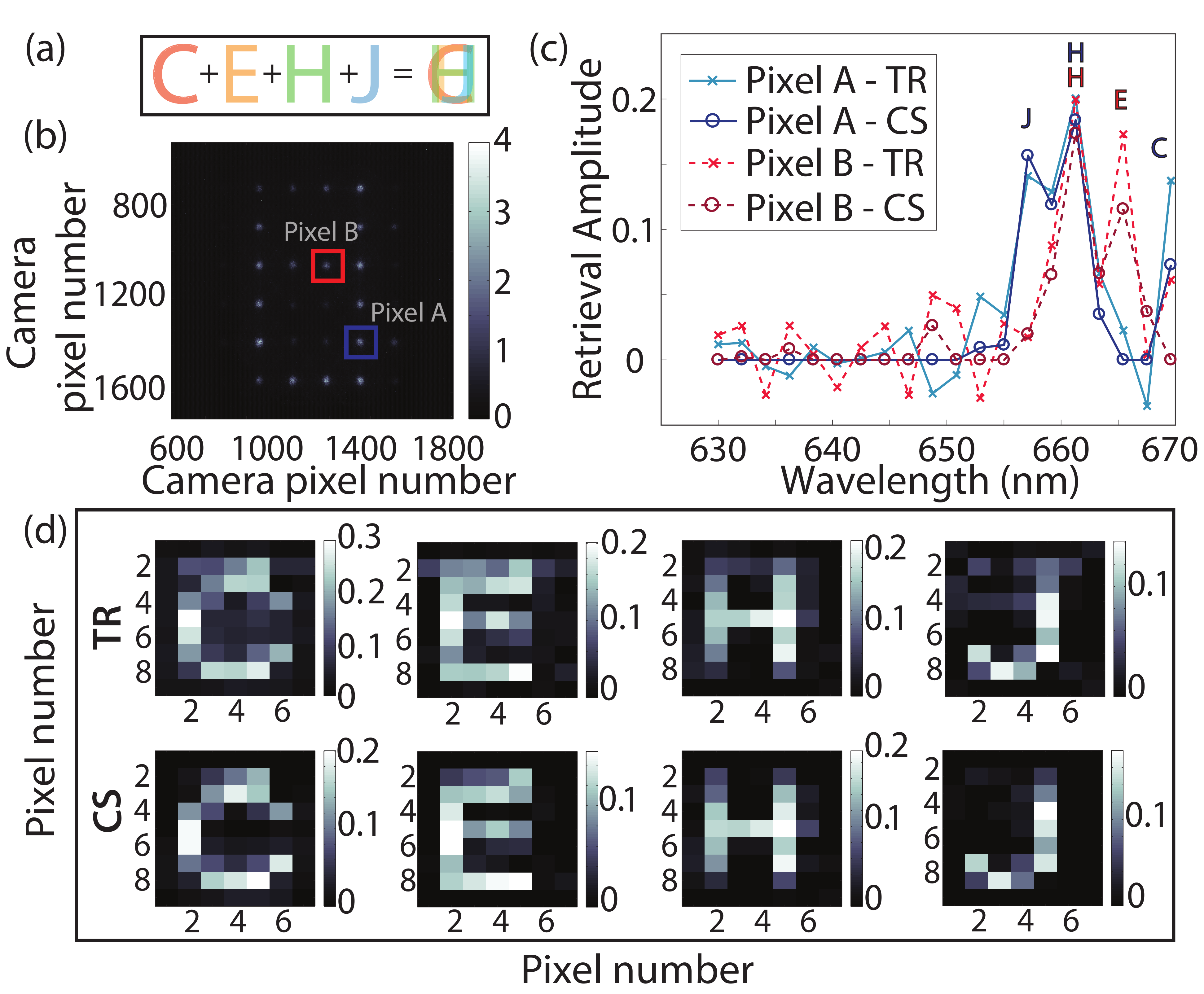}
		\caption{Reconstructing spatial and spectral information using TR and a CS technique. (a), (b) A composite image constructed of four experimental camera images (example shown in Figure~\ref{fig:Experiment}): an artist's impression of the experiment and raw data showing superimposed speckle patterns, respectively. (c) Wavelengths recovered from "Pixel A" and "Pixel B" in (b). (d) The reconstructed spatial intensities for each spectral channel.}
		\label{fig:Superimposedletters}
	\end{figure}

	Next, we compare the methods of TR and CS in an experimental spectral image reconstruction situation with multiple objects at different wavelengths, as illustrated in  Figure~\ref{fig:Superimposedletters} (a). Speckle images of the letter-shaped apertures taken at different wavelengths are superimposed numerically to produce a hyperspectral dataset as shown in Figure~\ref{fig:Superimposedletters} (b). To the human eye, the resulting monochromatic image does not appear to be made up of different shapes or wavelengths of light. Using the methods above, we reconstruct spectral information at each spatial coordinate in the image, yielding the results shown in  Figures~\ref{fig:Superimposedletters} (c) and (d). The illustrated reconstruction was done using $Y/X=25$, therefore both the TR and CS methods are appropriate methods to distinguish between frequencies and spatial coordinates in our imaging system. While both exhibit a high fidelity in terms of spectral and spatial reconstruction, it is clear that CS produces relatively low computational reconstruction noise when compared with TR. CS assumes that our input spectrum is sparse and, therefore, once peak values are determined, all others are assumed to be zero. This means that instead of our signal being degraded by computational noise, as in the TR method, we observe a clearly defined spectral and spatial reconstruction. Furthermore, we find that CS allows the recovery of a more complex spectrum than TR, due to its ability to suppress background computational noise.
	
	The use of transmission-matrix based methods for hyperspectral imaging has several specific advantages compared to conventional, dispersion based systems. The illumination of many camera pixels for each wavelength results in an increased dynamic range per wavelength for situations where spectra are reasonably sparse. Furthermore, as only a small area of each speckle contains information about all calibrated wavelengths, CS based techniques can be applied allowing for a reduction of the number of measurement points below the Nyquist-Shannon sampling limit. Our technique proves more advantageous than normal diffraction methods when dealing with measurements of sparse narrowband signals over a wide spectral range and within a small detection area. Alternatively, for an illumination system consisting of a discrete number of narrowband sources (i.e. lasers or LEDs) the system could be calibrated for only these wavelengths, thus allowing very efficient analysis of only these components. Such applications may include multiplexed imaging of narrowband light sources like LEDs in machine vision, or molecular fingerprinting techniques like Raman spectroscopy. The use of a thin nanowire mat offers an good optimum between transmission and spectral resolution, and has the potential of increasing the spatial resolution while maintaining low cross-talk between adjacent spatial coordinates. Further improvement of the stability and reproducibility of the scattering medium is a topic of ongoing study of interest for speckle-based imaging and encryption applications \cite{Goorden14}.
	
	While advantageous in specific applications, the scattering-based technique is clearly limited in spectral and spatial resolution. Ultimately, the transmission-matrix approach will break down by the self-averaging properties of speckles. The contrast of the speckle pattern is reduced proportional to the square root of the number of discrete spectral components. The successful reconstruction of up to 10 wavelengths in this study using CS techniques demonstrates that there is a window of opportunity for these techniques to be viable. The reconstruction is limited by the signal-to-noise ratio of the imaging system and the available photon budget, and faithful reconstruction of broadband signals with high spectral resolution has already been demonstrated for the case of multimode fiber systems \cite{Redding2014}. Application of this technique to narrowband features such as found in Raman spectroscopy remains challenging but appears within the range of possibilities. Perhaps one of the key challenges in this case is the maximization of throughput through the scattering medium, which involves the optimization of scattering strength as well as the collection efficiency of high-angle diffuse light after transmission.
	
	In conclusion, we have shown that hyperspectral imaging in the frequency domain can be achieved using a microlens array, a multiple scattering nanowire mat, and a monochromatic camera. The technique is based on the principle that we can characterize frequency-dependent speckle patterns at various spatial coordinates in a STM, and solve a linear system. We have compared two computational processes used to recover spectral and spatial information, and shown that CS can reconstruct a discrete spectrum to high precision.

	\section*{Funding Information}
	
	Defence Science \& Technology Laboratory (DSTL) (DSTLX1000092237). European Research Council (ERC) (278025).
	\section*{Acknowledgements}
	The authors thank Laurent Daudet for useful discussion. All data supporting this study are openly available from the University of Southampton repository at http://doi.org/10.5258/SOTON/D0006.


\end{document}